\documentclass{amsart}

\usepackage{amsmath,amsthm}
\usepackage{amsfonts,amssymb}

\newtheorem{thm}{Theorem}[section]
\newtheorem{cor}[thm]{Corollary}

\newtheorem{prop}[thm]{Proposition}

\theoremstyle{remark}

\def\a{{\alpha}}
\def\b{{\beta}}
\def\l{{\lambda}}
\def\t{{\theta}}
\def\g{{\gamma}}

 \def\CA{{\mathcal A}}
 
 \def\CD{{\mathcal D}}
 
 \def\CH{{\mathcal H}}

 \def\CO{{\mathcal O}}
 \def\CP{{\mathcal P}}
 \def\CR{{\mathcal R}}
 
 \def\CV{{\mathcal V}}

 \def\NN{{\mathbb N}}

 \def\RR{{\mathbb R}}

        \def\proj{\operatorname{proj}}
        \def\vol{\operatorname{vol}}

\include{graphicx}

\begin{document}

\title[Radon projections and orthogonal expansions]
{Reconstruction from Radon projections and orthogonal expansion on a ball}

\author{Yuan Xu}
\address{Department of Mathematics University of Oregon
    Eugene, Oregon 97403-1222.}
  \email{yuan@math.uoregon.edu}

\date{\today}
\thanks{ The author  was partially supported
by the National Science Foundation under Grant DMS-0604056}

\begin{abstract}
The relation between Radon transform and orthogonal expansions of 
a function on the unit ball in $\RR^d$ is exploited. A compact formula 
for the partial sums of the expansion is given in terms of the Radon
transform, which leads to algorithms for image reconstruction from 
Radon data. The relation between orthogonal expansion and the 
singular value decomposition of the Radon transform is also exploited.  
\end{abstract}

\maketitle

\section{Introduction}
\setcounter{equation}{0}

Reconstruction of an image from its Radon projections is the 
central theme in x-ray tomography and has spectacular 
applications in medical imaging. Mathematically the problem 
is to find a good approximation to a function based on a finite
collection of its Radon projections (see, for example, 
\cite{He,KS,N}).

The main topic of this paper is the connection between the 
Radon transform and the orthogonal expansion of the function
on a unit ball. This connection was initiated in the classical 
paper \cite{Co} with an inversion formula of the Radon transform 
based on spherical harmonic expansions. The relation between 
the Radon transform of a function, supported on the unit ball, 
and its orthogonal expansion was studied or used in 
\cite{Da,DaG,LS,Lo, Ma, Marr}, among others (see \cite{N} for 
further references). The papers \cite{Da, DaG, Lo} studied also the 
singular value decomposition (SVD) of the Radon transform 
using an orthogonal basis. Since then SVD has become an 
important tool for studying the stability of the inversion 
problem, the resolution of the reconstruction, and the
incomplete data problem; see, for example, \cite{CB, DaG, 
Lo2, Maa, N}. The truncated SVD also provides an algorithm
for reconstruction of images. Because of the complicated 
formulas involved in the orthogonal or SVD expansions (see, 
for example, \cite{Da,Lo,N}), the algorithms did not seem to 
be used in practical applications.

Recently a new reconstruction algorithm was proposed in 
\cite{X05} and further studied in \cite{XT,XTC}. The new 
algorithm is called OPED, as it is based on orthogonal 
polynomial expansion on the unit disk. The algorithm
reproduces polynomials of high degrees and allows a fast 
implementation (\cite{XT}). The numerical tests shows 
that the algorithm is fast, stable, and produces high quality
images (\cite{XT, XTC}. The key ingredient  for deriving the 
algorithm is the following formula for the  partial sum 
$S_{2m} f$ of the orthogonal expansion of $f$ on the 
unit disk, 
\begin{equation}\label{1stS2m}
   S_{2m} f(x,y) = \frac{1}{2m+1}\sum_{\nu=0}^{2m} 
       \int_{-1}^1 \CR_{\phi_\nu} f(t)
  \Phi_{2m}(t, x \cos \phi_\nu+y \sin \phi_\nu)dt,
\end{equation}
where  $\phi_\nu = \frac{2 \nu \pi}{2m+1}$ and 
$\CR_\theta f(t)$ is the Radon projection on the line 
$x \cos \theta +  y \sin \theta = t$ (see Section 3). It turns
out that there is a natural extension of this formula to the 
unit ball of higher dimension, which shows that the 
orthogonal polynomial expansion of $f$ can be expressed 
in terms of the Radon transforms and allows us to extend
the OPED algorithm in the unit ball of $\RR^d$.  Furthermore, 
there is a close relation between SVD and the extension of 
the formula \eqref{1stS2m}. In fact, they can be brought 
together by the use of a compact formula of the reproducing 
kernel of orthogonal polynomials in \cite{X99}.  The 
orthogonal expansion on the unit ball has been studied 
recently in \cite{X99,X05a},  which can be used, in particular, 
to derive the uniform convergence of the algorithms. 

The purpose of this paper is two folds. Firstly we will
clarify the relation between orthogonal expansion on the ball
and the Radon projections and derive the extension of the 
OPED algorithm in higher dimensions. Secondly, we will 
explain the connection between SVD of the Radon transform
and orthogonal expansions. In particular, we shall show that 
using truncated SVD to reconstruct the image is the same 
as using OPED algorithm. 

The paper is organized as follows. The following section 
contains a succinct account of the basic results on 
orthogonal polynomials on the unit ball. The orthogonal 
expansions in terms of the Radon projections is developed 
in Section 3. The extension of the OPED algorithms and 
a convergence result are given in Section 4. Finally, the 
SVD of the Radon transform is discussed in Section 5. 

\section{Preliminaries on orthogonal polynomials}
\setcounter{equation}{0}

Let $B^d:= \{x: \|x\| \le 1\}$ and $S^{d-1}:=\{x:\|x\| =1\}$ be
the unit ball and the unit sphere of $\RR^d$, respectively. We 
denote the surface area of $S^{d-1}$ by $\sigma_d$ and the 
volume of $B^d$ by $b_d$. Then
$$
    \sigma_d = \frac{2 \pi^{d/2}}{\Gamma(d/2)} 
         \qquad \hbox{and} \qquad 
    b_d = \frac{\sigma_d}{d} = \frac{\pi^{d/2}}{ \Gamma((d+2)/2)}.
$$

{\it Inner product on the ball.} 
For later discussion let us introduce a weight function $W_\mu$ on 
the unit ball, 
$$
    W_\mu(x) = (1-\|x\|^2)^{\mu -1/2}, \qquad x\in B^d. 
$$
The inner product on the unit ball is denoted by 
$$
 \langle f, g \rangle_{L^2(B^d)}  = 
      a_{\mu} \int_{B^d} f(x) g(x) W_\mu(x) dx 
$$
where $a_{\mu}$ is the normalization constant of $W_\mu$, 
that is, $a_{\mu} = 1/\int_{B^d} W_\mu(x) dx$. For $\mu =1/2$, 
it is the unit weight (Lebesgue measure) and $a_\mu$ is equal to 
$b_d^{-1}$. We will mainly work with the Lebesgue measure, so 
the inner product $\langle f, g \rangle_{L^2(B^d)}$ should be 
regarded as with $\mu = 1/2$ unless specified otherwise.

\medskip

{\it Polynomial spaces.} Let $\Pi_n^d$ denote the space of 
polynomials of degree $n$ in 
$d$ variables. We say that $P \in \Pi_n^d$ is an orthogonal 
polynomial on $B^d$ if $\langle P, Q \rangle_{L^2(B^d)} =0$ for 
all $Q \in \Pi_{n-1}^d$. Let $\CV_n^d$ denote the space of 
orthogonal polynomials. It is well-known that 
$$
  \dim \Pi_n^d = \binom{n+d}{n} \qquad\hbox{and}\qquad
  \dim \CV_n^d =\binom{n+d-1}{n}.
$$
Several explicit orthonormal bases of $\CV_n^d$ are known
(see, for example, \cite{DX}). We will need one given in terms
of the Jacobi polynomials and spherical harmonics.

\medskip

{\it Jacobi polynomials.} The $k$-th Jacobi polynomial is 
dentoed by $P_k^{(\a,\b)}$ and they satisfy the orthogonal 
relation (\cite{Sz})
\begin{align}\label{Jacobi}
&  c_{\a,\b} \int_{-1}^1 P_k^{(\a,\b)}(t)P_l^{(\a,\b)}(t) w_{\a,\b}(t)dt \\
& \qquad\qquad  = 
   \frac{(\a+1)_k(\b+1)_k(\a+\b+k+1)}{k!(\a+\b+2)_k(\a+\b+2k+1)} \delta_{k,l}
   :=h_k^{(\a,\b)} \delta_{k,l}, \notag
\end{align}
where $w_{\a,\b}(t)=  (1-t)^\a(1+t)^\b$,  $c_{\a,\b}$ is the 
normalization constant of $w_{\a,\b}$, 
$$
   [ c_{\a,\b} ]^{-1} = \int_{-1}^1 w_{\a,\b}(t) dt = 2^{\a+\b+1}
        \frac{\Gamma(\a+1)\Gamma(\b+1)}{\Gamma(\a+\b+2)},
$$ 
and the notation $(a)_k : = a (a+1) \cdots (a+k-1)$ denotes the 
shifted factorial (Pochhammer symbol).  From \eqref{Jacobi} 
the orthonormal Jacobi polynomials are given by $
p_n^{(\a,\b)}(t):= [h_n^{(\a,\b)}]^{-1/2} P_n^{(\a,\b)}(t)$. 

\medskip

{\it Gegenbauer polynomials and Chebyshev polynomials.}
When $\alpha = \beta = \lambda -1/2$, the Jacobi polynomials
become the Gegenbauer polynomials, usually denoted by 
$C_k^\lambda$ and normalized by 
\begin{equation} \label{Gegen}
 c_\lambda \int_{-1}^1 C_k^{\l}(t)C_l^{\l}(t) (1-t^2)^{\l-1/2}dt 
   = \frac{\l (2\l)_k}{(k+\l)k!} \delta_{k,l}: = h_k^{(\lambda)} \delta_{k,l}. 
\end{equation}
where $c_\l = \Gamma(1/2) \Gamma(\l + 1/2)/ \Gamma(\l +1)$.
When $\l =1$ and $\l=0$, $C_k^\l(t)$ becomes the Chebyshev 
polynomial of the second kind, $U_k(t)$, and the first kind, 
$T_k(t)$, respectively, and 
\begin{equation} \label{cheby}
    U_k(t) = \frac{\sin (k+1)\t}{\sin \t} \qquad \hbox{and} \qquad 
     T_k(t) = \cos k \t, \quad\hbox{where}\quad t = \cos \t. 
\end{equation}

\medskip

{\it Spherical harmonics.} These are defined as the restriction 
of the homogeneous harmonic polynomials on the sphere. Let 
$\CH_n^d$ denote the space of spherical harmonics of degree 
$n$ in $d$ variables. It is known that 
$$
  \dim \CH_n^d = \binom{n+d-1}{n}  - \binom{n+d-3}{n}. 
$$
Let $\{Y_{k,n}: 1 \le k \le \dim \CH_n^d\}$ denote an orthonormal
basis of $\CH_n^d$. Then 
$$
 \sigma_d^{-1} \int_{S^{d-1}} Y_{k,n}(\xi) Y_{l,n}(\xi)d\omega(\xi) 
      = \delta_{k,l}, \qquad 1 \le k,l \le \dim\CH_n^d.
$$
We emphasis that $Y_{k,n}(x)$ are in fact homogeneous 
polynomials in $\Pi_n^d$. 

\medskip

{\it An orthonormal basis for $\CV_n^d$.} We give the basis for 
inner product defined in terms of $W_\mu(x)$. Setting $\mu =1/2$
gives the basis for the Lebesgue measure. Let $Y_{j,m}$ be
an orthonormal basis for $\CH_m^d$. Define
\begin{equation} \label{OPbasis}
f_{k,j}^n(x) = [h_{n,k}]^{-1} 
  p_k^{(\mu -\frac{1}{2},n-2k+\frac{d-2}{2})}(2\|x\|^2 -1) 
         Y_{j,n-2k}(x), 
\end{equation}
where 
$$
   [h_{n,k}]^2 :=\frac{\Gamma(\mu + \frac{d+1}{2})
      \Gamma(n-2k + \frac{d}{2})}{\Gamma(\frac{d}{2})
        \Gamma(n-2k+\mu+\frac{d+1}{2})}. 
$$
Then the set $\{f_{k,j}^n: 1 \le j \le \dim \CH_{n-2k}^d, 
  0 \le 2 k \le n\}$ is an orthonormal basis for $\CV_n^d$;
that is, $f_{k,j}^n \in \CV_n^d$ and 
$ \langle f_{k,j}^n, f_{k',j'}^{n} \rangle_{L^2(B^d)} = 
       \delta_{k,k'}\delta_{j,j'}$ (see \cite[p. 39]{DX}).
  
\medskip

{\it Reproducing kernel of $\CV_n^d$}. The reproducing kernel
$P_n(\cdot,\cdot)$ of $\CV_n^d$ satisfies
\begin{equation}\label{reprod0}
   a_\mu \int_{B^d} f(y) P_n(x,y) W_\mu(y)dy = f(x),
        \qquad \forall f\in \CV_n^d.
\end{equation}
Let $\{P_k^n:  1 \le k \le \dim \CV_n^d\}$ denote any orthonormal 
basis of $\CV_n^d$.  Then
$$
P_n(x,y) = \sum_{k=1}^{N_n} P_k^n(x) P_k^n(y), \qquad 
           N_n = \dim \CV_n^d. 
$$
The definition of $P_n(\cdot, \cdot)$, however, is independent 
of the particular choice of bases. In particular, we can take the 
orthonormal basis in \eqref{reprod} and get 
\begin{equation}\label{reprod}
 P_n(x,y) = \sum_{0 \le 2k \le n} \sum_{j=1}^{\dim \CH_{n-2k}^d}
       f_{k,j}^n(x) f_{k,j}^n(y).
\end{equation}
The reproducing kernel satisfies a compact formula that will 
play a fundamental role in our  study; it is given by (\cite{X99})
$$
 P_n(x,y) = \frac{n+\l}{\l} c_{\mu-\frac{1}{2}}
   \int_{-1}^1C_n^\l (\langle x, y\rangle + 
      \sqrt{1-\|x\|^2}\sqrt{1-\|y\|^2} \,s) (1-s^2)^{\mu-1} ds
$$
where $\lambda = \mu + \frac{d-1}{2}$, $\langle \cdot, \cdot
\rangle$ is the Euclidean inner product in $\RR^d$, and 
$c_\lambda$ is defined in \eqref{Gegen}. In particular, it 
implies that
\begin{equation}\label{reprod2}
    P_n(x,\xi) = \frac{n+\l}{\l} C_n^\l (\langle x, \xi\rangle),
         \qquad \xi \in S^{d-1}, \quad x \in B^d. 
\end{equation}

\medskip

{\it Orthogonal expansions on $B^d$.} If $\{P_k^n: 1 \le k 
\le N_n\}$, $N_n=\dim \CV_n^d$, is an orthonormal basis of 
$\CV_n^d$, then the standard Hilbert space theory states that 
there is an orthogonal expansion
$$
f  = \sum_{k=0}^\infty 
  \sum_{k=1}^{N_n} \langle f, P_k^n\rangle_{L^2(B^d)}       
   P_k^n, \qquad \forall f \in L^2(B^d). 
$$ 
Let $\proj_k: L^2(B^d) \mapsto \CV_n^d$ denote the projection
operator. Using the reproducing kernel,  the orthogonal expansion 
can be stated as 
\begin{equation} \label{OrthExpand}
   f = \sum_{k=0}^\infty \proj_k f, \qquad \hbox{where} \quad
        \proj_k f = a_\mu \int_{B^d} f(y) P_n(x,y) W_\mu(y) dy,
\end{equation}
which is independent of the particular choices of the bases of 
$\CV_n^d$. 


\section{Radon Transform and Orthogonal Polynomial Expansion}
\setcounter{equation}{0}

Let $f \in L^1$ be a real valued function. For $\xi \in S^{d-1}$ 
and $t \in \RR$, the Radon transform of $f$ is defined as
$$
    \CR f(\xi, t) := \int_{\langle \xi, x\rangle = t} f(x) dx
         = \int_{\xi^\perp} f (t \xi + y )dy,
$$
where the integral is over a hyperplane of $(d-1)$-dimension 
perpendicular to $\xi$ and with minimum distance $t$ to the origin. 
More general definition on other spaces or manifolds can be found
in \cite{Hel}. For properties of Radon transforms we refer to 
\cite{Hel,N}.  We assume 
that $f$ has compact support in $B^d$, so that the
integral above should be understood as over $B^d \cap 
\{x: \langle \xi, x\rangle = t\}$. In particular, for $\xi \in S^{d-1}$,
let $Q_\xi$ denote an orthogonal matrix whose first row is $\xi$
and let $B^d(r)$ denote the ball of radius $r$ in $\RR^d$; then 
a change of variables $x \mapsto (t, y)Q_\xi$ shows that 
\begin{align} \label{Radon}
 \CR f(\xi,t) & = \int_{B^{d-1}(\sqrt{1-t^2})} f((t, y) Q_\xi)dy \\
 & = (1-t^2)^{\frac{d-1}{2}} \int_{B^{d-1}} f((t, \sqrt{1-t^2} y) Q_\xi)dy. 
  \notag
\end{align}
Since $\langle (t,y)Q_\xi, \xi \rangle = t$, an immediate consequence 
of \eqref{Radon} is the following identity, 
\begin{equation} \label{IntRf}
   \int_{B^d} f(x) g(\langle x,\xi \rangle) dx = 
        \int_{-1}^1 \CR f(\xi,t) g(t) dt, \qquad \xi \in S^{d-1},
\end{equation}
whenever both integrals make sense. The definition of $\CR f$ also 
implies that 
\begin{equation} \label{eq:RfEven}
     \CR f(- \xi, -t) = \CR f(\xi, t), \qquad \xi \in S^{d-1}, \quad t\in \RR.
\end{equation}
For fixed $\xi$ and $t$, we also call $\CR f(\xi,t)$ a Radon 
projection.  The essential problem for x-ray imaging is to find a 
good approximation to the function $f$ based on a given data 
set of its Rdaon projections. 

We now derive the orthogonal expansion of $f$ in terms of 
Radon projections. The following proposition plays a key role. 

\begin{prop} \label{Reprod}
For $x,y \in B^d$, the reproducing kernel $P_n(\cdot,\cdot)$ 
satisfies
\begin{equation}\label{r_kernel}
    P_n(x,y) = \frac{n+d/2}{d/2}  \sigma_d^{-1} \int_{S^{d-1}}
         C_n^{d/2}(\langle x,\xi\rangle) C_n^{d/2}(\langle y,\xi\rangle)
          d\omega(\xi). 
\end{equation}
\end{prop}

\begin{proof}
From the explicit formula of $f_{k,j}^n$ at \eqref{OPbasis} with 
$\mu =1/2$, we deduce that 
\begin{equation} \label{f_xi}
      f_{k,j}^n(\xi) = H_n Y_{j,n-2k}(\xi), \qquad \xi \in S^{d-1},   
\end{equation}
where, using the fact that $p_k^{(0,\b)}(t) = [h_k^{(0,\b)}]^{-1/2} 
P_k^{(0,\b)}(t)$, $P_k^{(0,\b)}(1) = 1$, and the formula of 
$h_k^{(\a,\b)}$ in \eqref{Jacobi}, we have
\begin{equation} \label{Hk}
 H_n  =  [h_{n,k}]^{-1} 
  p_k^{(0,n-2k+\frac{d-2}{2})}(1) = \sqrt{\frac{n+d/2}{d/2}},
\end{equation}
independent of $k$. Consequently, integrating over $S^{d-1}$
we get
$$
   \sigma_{d}^{-1} \int_{S^{d-1}} f_{k,j}^n(\xi) f_{k',j'}^{n}(\xi) 
       d\omega(\xi) = H_n^2 \delta_{j,j'}\delta_{k,k'}
         = \frac{n+d/2}{d/2} \delta_{j,j'}\delta_{k,k'}.
$$
Multiplying the above equation by $f_{k,j}^n(x)$ and 
$f_{k',j'}^n(y)$ and summing over all $j, j', k, k'$, the stated
result follows from \eqref{reprod} and \eqref{reprod2}.
\end{proof}

\begin{thm} \label{thm:3.2}
For $n \ge 0$,
$$
  \proj_n f(x) = \frac{n+d/2}{d/2} \sigma_d^{-1} \int_{S^{d-1}}
       b_d^{-1} \int_{-1}^1 \CR f(\xi,t) C_n^{d/2} (t) dt \,
            C_n^{d/2}(\langle x,\xi\rangle) d \omega(\xi). 
$$
In particular, for $f \in L^2(B^d)$, 
\begin{equation} \label{expand}
   f = \sum_{n=0}^\infty \frac{n+d/2}{d/2} \sigma_d^{-1} \int_{S^{d-1}}
       b_d^{-1} \int_{-1}^1 \CR f(\xi,t) C_n^{d/2} (t) dt \,
            C_n^{d/2}(\langle \cdot,\xi\rangle) d \omega(\xi). 
\end{equation}
\end{thm}

\begin{proof}
By the formula \eqref{reprod0} with $\mu =1/2$ and the 
formula \eqref{r_kernel} of $P_n(\cdot,\cdot)$ we have 
\begin{align*}
     \proj_n f(x) & =  b_d^{-1} \int_{B^d} f(y) P_n(x,y) dy \\
   &   =       \frac{n+d/2}{d/2}  \sigma_d^{-1} \int_{S^{d-1}}
     b_d^{-1} \int_{B^d} f(y) C_n^{d/2}(\langle y,\xi\rangle) dy
      C_n^{d/2}(\langle x,\xi\rangle)   d\omega(\xi).  
\end{align*}
The identity \eqref{IntRf} shows that the inner integral is
\begin{equation} \label{eq:Int1}
    \int_{B^d} f(y)C_n^{d/2}(\langle y,\xi\rangle) dy 
     = \int_{-1}^1 \CR f(\xi,t) C_n^{d/2} (t) dt, 
\end{equation}
so that the stated formula follows. 
\end{proof}

The formula \eqref{expand} as stated here has already appeared 
in \cite{P} in a study of the approximation by ridge functions. See
also \cite{BG} for the case of $d =2$.  Although spherical harmonics
expansions for $d=2$ was used in the classical work of \cite{Co}, its 
compact form in \eqref{expand} is quite recent and not used for 
reconstructing images from Radon data until recently (\cite{X05}). 
It should also be noted that for $d > 2$, the Gegenbauer polynomials
and spherical harmonics were used for constructing Radon transforms 
already in \cite{Lo}. 

Let us mention that there does not seem to be an analogous 
formula for the more general case of orthogonal expansion with
respect to $W_\mu$. In fact, in the general case, the formula 
\eqref{OPbasis} gives 
$$
   f_{k,j}^n(\xi) = H_{n,k} Y_{j,n-2k}(\xi),  \qquad \xi \in S^{d-1},
$$
where 
\begin{equation} \label{Hnk}
     H_{n,k} := \frac{(\mu+1/2)_k (\mu+\frac{d-1}{2})_{n-k}
          (n+\mu+\frac{d-1}{2}) }
                   {k! (\frac{d}{2})_{n-k} (\mu + \frac{d-1}{2})},
\end{equation}
which depends on both $n$ and $k$ (comparing with \eqref{Hk}),
so that Proposition \ref{Reprod} with $C_n^{d/2}$ replaced by
$C_n^{\mu + \frac{d-1}{2}}$ does not hold. 

Let $S_n f$ denote the partial sum operator of the 
orthogonal expansion \eqref{OrthExpand}, 
\begin{equation} \label{Snf}
      S_n f(x) = \sum_{k=0}^n \proj_k f(x).  
\end{equation}  
Evidently, the expansion \eqref{OrthExpand} holds in the sense
that $S_n f \to f$ in $L^2(B^d)$ norm.  

\begin{cor}
Let $S_n$ be the partial sum operator defined in \eqref{Snf}.
Then 
\begin{equation} \label{eq:Sn}
 S_n f(x) = \sigma_d^{-1} \int_{S^{d-1}}b_d^{-1} 
 \int_{-1}^1 \CR f(\xi,t) \Phi_n( t, \langle x,\xi\rangle) dt d \omega(\xi). 
\end{equation}
where 
\begin{equation} \label{eq:Phi}
   \Phi_n(t, u) : = \sum_{k=0}^n \frac{k+d/2}{d/2} 
            C_k^{d/2} (t) C_k^{d/2}(u).  
\end{equation}
\end{cor}

A cubature formula on $S^{d-1}$ of degree $M$ is a discrete 
sum such that 
\begin{equation} \label{cubature}
   \sigma_d^{-1}  \int_{S^{d-1}} f(\xi) d\omega(\xi) = 
     \sum_{\nu=1}^N \lambda_\nu f(\xi_\nu), 
        \qquad f \in \Pi_M(S^{d-1}), 
\end{equation}
where $\Pi_M(S^{d-1})$ is the space of spherical polynomials,
that is, the space of $\Pi_M^d$ restricted on $S^{d-1}$. If all 
$\lambda_k$ are positive, the cubature is called {\it positive}.  
We call a polynomial $P \in \Pi_M^d$ {\it even} if it satisfies 
$P(x) = P(-x)$ for all $x \in \RR^d$. The cubature formula 
\eqref{cubature} is called {\it symmetric}, if it is exact for all 
even polynomials in $\Pi_M(S^{d-1})$. 

\begin{prop}
Suppose \eqref{cubature}  is a symmetric cubature formula of 
degree $2 n$. Then 
\begin{equation} \label{SnSemiDis}
   S_n f(x) = \sum_{\nu=1}^{N} \lambda_\nu b_d^{-1} 
    \int_{-1}^1 \CR f(\xi_\nu,t) \Phi_n(t, \langle x,\xi_\nu\rangle ) dt. 
\end{equation}
\end{prop}

\begin{proof}
The equation \eqref{eq:Int1} shows that $P_x(\xi):=
\int_{-1}^1 \CR f(\xi,t) \Phi_n(\xi,t;x)dt $ is a 
polynomial of degree at most $2n$ in $\xi$. Furthermore, 
using the fact that $\CR f(-\xi,-t) = \CR f(\xi,t)$, it is easy to see 
that $P_x$ is even, so that the cubature formula on $S^{d-1}$
is exact when applied to $P_x(\xi)$.
\end{proof}

We consider some special cases of lower dimensions below. 

\medskip\noindent
{\bf The case d=2.}  For $\xi \in S^1$ we write $\xi = (\cos \t, \sin \t)$
and we shall write $\CR_\t f(t)$, $\t \in [0,2\pi]$, instead of 
$\CR f(\xi,t)$. Since $b_2 = \pi$ and the following cubature formula
$$
 \frac{1}{2\pi}\int_{S^1} f(\xi) d\omega(\xi) = \frac{1}{n+1} 
     \sum_{\nu =0}^n f(\xi_\nu),
 \qquad \xi_\nu = (\cos \tfrac{\nu \pi}{n+1}, \sin \tfrac{\nu \pi}{n+1})
$$
is symmetric and of degree $2n$, we conclude that 
\begin{equation}\label{d2Sn}
   S_n f(x) = \frac{1}{n+1}\sum_{\nu=0}^{n} 
    \int_{-1}^1 \CR_{\t_\nu} f(t)
  \Phi_n(t, x_1 \cos \t_\nu+x_2 \sin \t_\nu)dt 
\end{equation}
where $\t_\nu = \tfrac{\nu\pi}{n+1}$ and $\Phi_n$ is \eqref{eq:Phi} for
$d =2$,
$$
   \Phi_n(t, u) = \sum_{k=0}^n (k+1) U_k(t) U_k(u).
$$
This formula can be found implicitly in \cite{LS} (see (5.9), (4.3) 
and (3.7) there).
In the case of $n = 2m$, we can use the elementary relations 
$$
\cos \tfrac{(2\nu + 1)\pi}{2m+1}  = - \cos \tfrac{(2(\nu+m) \pi}{2m+1}, 
\quad \qquad
\sin \tfrac{(2\nu + 1)\pi}{2m+1}  = - \sin \tfrac{(2(\nu+m) \pi}{2m+1} 
$$
and the fact that $\CR(\t + \pi, -t) = \CR(\t, t)$ to rewrite 
\eqref{d2Sn} as
\begin{equation}\label{d2Sn2}
   S_{2m} f(x) = \frac{1}{2m+1}\sum_{\nu=0}^{2m} 
       \int_{-1}^1 \CR_{\phi_\nu} f(t)
  \Phi_{2m}(t, x_1 \cos \phi_\nu+x_2 \sin \phi_\nu)dt
\end{equation}
where $\phi_\nu =\tfrac{2\nu\pi}{2m+1}$. This is the formula 
\eqref{1stS2m} proved in \cite{X05} from which the OPED 
algorithms are derived.  
\qed

\medskip\noindent
{\bf The case d=3.}  For $\xi \in S^2$ we use the spherical 
coordinate
$$
\xi = (\sin \phi \sin \t, \sin \phi \cos \t, \cos \phi), \qquad 
      0\le \phi \le \pi, 0 \le \t \le 2 \pi.
$$
Several explicit cubature formulas on the sphere are known, 
see, for example, \cite{My, St}. Let $t_{k} = \cos \t_k$,
$k =0, 1, \ldots,n$, denote the zeros of the Legendre polynomial
of degree $n+1$ and $\lambda_{k}$ be the corresponding 
weights of the Legendre-Gaussian quadrature formula. Let 
$$
\xi_{k,\nu} = (\sin \tfrac{\nu \pi}{n+1} \sin \t_k, 
   \cos \tfrac{\nu}{n+1} \sin \t_k, \cos \t_k), \qquad 0 \le k, \nu \le n 
$$
Then the product type cubature formula 
$$
  \frac{1}{4 \pi} \int_{S^2} f(\xi)d\omega(\xi) = \frac{1}{n+1}
       \sum_{k=0}^n \lambda_k \sum_{\nu=0}^n
          f(\xi_{k,\nu})  
$$
is symmetric and of degree $2n$. Consequently, we have 
\begin{equation}\label{d3Sn}
   S_n f(x) = \frac{1}{n+1} \sum_{k=0}^n \lambda_k
     \sum_{\nu=0}^{n} \int_{-1}^1 \CR f (\xi_{k,\nu}, t)
     \Phi_n(t, \langle  \xi_{k,\nu},x\rangle) dt,  
\end{equation}
where $\Phi_n$ is the function \eqref{eq:Phi} for $d=3$.  \qed 
\medskip
 
The formula of $S_n f$ in terms of Radon projections allows 
us to give an approximation to $f$ based on finite Radon 
projections. The convergence of $S_n f$ to $f$ holds in $L^2$
norm but does not hold in the uniform norm in general. In fact, 
it is known that \cite{X01}
\begin{equation}\label{SnNorm}
    \|S_n\|_\infty = \CO(n^{\frac{d-1}{2}}), \qquad d \ge 2,  
\end{equation}
where $\|\cdot\|_\infty$ is the operator norm of $S_n$ in $C(B^d)$,
and $A_n = \CO(B_n)$ means $c_1 A_n \le B_n \le c_2 A_n$ for 
two constants $c_1$ and $c_2$ independent of $n$. There is,
however, a simple construction that gives a better convergence 
result. 

Let $\eta$ be a $C^{d+2}(\RR)$ function such that $\eta(t) \ge 0$,
$\eta(t) =1$ for $0 \le t \le 1$ and $\eta$ has compact support on 
$[0,2]$. Define 
\begin{equation}\label{Sn_eta}
  S_n^\eta f (x) : = \sum_{k=0}^{2n} 
      \eta \left(\frac{k}{n}\right) \proj_k f(x). 
\end{equation} 
The operator $S_n^\eta$ satisfies the following properties 
\cite{X05a}: 

\begin{prop} Let $f \in L^p(B^d)$, $1 \le p < \infty$ or $f \in C(B^d)$
for $p = \infty$. Then
\begin{enumerate}
\item  $S_n^\eta f  =  f$ if $f \in \Pi_n$;   
\item  for $ n \in \NN$, $\| \eta_nf \|_p \le c \|f\|_p$
\item  for $ n \in \NN$, 
$ 
    \|f - \eta_n \|_p \le c E_n(f)_p : = \inf_{p \in \Pi_n^d} \|f - p\|_p.
$
\end{enumerate}
\end{prop}

As $S_n^\eta f$ is a polynomial of degree $2n$, the last property
shows that, up to a constant multiple, it is close to the polynomial
of the best approximation to $f$. Since $\proj_n f$ can be written in 
terms of Radon projections, so can $S_n^\eta f$.


\section{OPED algorithms for reconstruction of images}
\setcounter{equation}{0}

The essential problem in computerized tomography is to 
find a good approximation to the function $f$ based on a
set of discrete Radon data. The expression \eqref{SnSemiDis} 
allows us to derive such an approximation by a simple 
quadrature formula on $[-1,1]$. Because of \eqref{Radon},
we choose the quadrature formula to be of the form
\begin{equation} \label{quadrat}
   c_{d/2}  \int_{-1}^1 f(t) (1-t^2)^{\frac{d-1}{2}} dt = 
      \sum_{j=0}^n w_j f(t_j),
\end{equation}
where $c_{d/2}$ is defined as in \eqref{Gegen},  and assume 
that it is exact for polynomials of degree $M$. In particular, 
we can choose the Gaussian quadrature, for which
$t_j= t_{j,n}$, $0\le j \le n$, are zeros of the Gegenbauer 
polynomial $C_{n+1}^{d/2}(t)$ and $w_j$ are all positive
and given by explicit formula (see \cite{Sz}). The Gaussian
quadrature formula is exact for polynomials of degree up
to $2n +1$. 

\begin{prop}
Let \eqref{cubature} be a positive symmetric cubature formula 
of degree $2n$ and \eqref{quadrat} be the Gaussian quadrature 
formula. Define
\begin{equation} \label{CAn}
   \CA_n f(x) = b_d^{-1} \sum_{\nu =1}^N \lambda_\mu
       \sum_{j=0}^n w_j  \CR f(\xi_\nu, t_j) 
         \Phi_n(t_j,\langle x,\xi_\nu\rangle). 
\end{equation}
Then $\CA_n f$ preserves polynomials of degree $n$, that is,
$\CA_n f = f$ whenever $f \in \CP_n^d$. 
\end{prop}

\begin{proof}
We start from \eqref{SnSemiDis}. If $f$ is a polynomial of 
degree at most $n$ then, by \eqref{Radon}, 
$(1-t^2)^{- \frac{d-1}{2}} \CR f(\xi_\nu, t)$ is a polynomial 
of degree $n$. As $\Phi_n (t, \langle x,\xi_\nu\rangle)$ is 
a polynomial of degree $n$ in $t$ and the Gaussian quadrature
formula is of degree $2n+1$, the fact that $\CA_n f = f$ 
follows.  
\end{proof}

The functions $\CA_n f$ are obtained from the orthogonal partial
sums $S_n f$ of $f$ by applying the Gaussian quadrature formula. They provide a sequence of approximation to $f$ based on the 
set of discrete Radon data
$$
  \{ \CR f(\xi_\nu, t_j): \quad 1 \le \nu \le N, 0 \le j \le n\}.
$$
In other word, $\CA_n$ provides an algorithm for reconstruction 
of images from the Radon data. We will show that $\CA_n f$ 
converges to $f$ uniformly if $f$ is smooth enough. First, however,
we consider some special cases. 

\medskip\noindent
{\bf The case d=2.} In this case we can start from the formula 
of $S_{2m}$ at \eqref{d2Sn2}. The Gaussian quadrature 
formula is 
$$
  \frac{1}{\pi} \int_{-1}^1 f(t) \sqrt{1-t^2} dt = \frac{1}{2m+1}
      \sum_{j=1}^{2m} \sin^2 \psi_j f(\cos \t_j), 
          \qquad \t_j =  \frac{j \pi}{2m+1},
$$ 
which leads to the OPED algorithm of type II,
\begin{equation} \label{2AnU}
   \CA_{2m} f(x) = \sum_{\nu =0}^{2m} \sum_{j=1}^{2m}
       \CR_{\phi_\nu} f(\cos \t_j) T_{j,\nu}(x), 
\end{equation}
where
$$
 T_{j,\nu}(x) = \frac{1}{(2m+1)^2} \sum_{k=0}^{2m} (k+1) 
   \sin ((k+1) \t_j) U_k(x_1 \cos \phi_\nu + x_2 \sin \phi_\nu).
$$
The OPED of type II is closely related to an algorithm in \cite{BO},
where the connection to orthogonal polynomial expansion was 
not considered. The formation of the lines on which the Radon 
projections  take place is often refereed to as scanning geometry, 
as it determines how the object being examined is scanned by 
the x-rays.  We can use the Gaussian quadrature formula for the 
Chebyshev polynomials of the first kind, 
$$
 \frac{1}{\pi} \int_{-1}^1f(t) \frac{dt}{\sqrt{1-t^2}} = \frac{1}{2m+1}
      \sum_{k=0}^{2m} f(\cos \psi_j), \qquad 
         \psi_j = \frac{(j+\frac{1}{2})\pi}{2m+1}, 
$$
to discretize the integral in \eqref{d2Sn2} by applying it to the
integrant multiplied by $1-t^2$,  leading to the OPED
algorithm of type I with a different scanning geometry, which 
has the same formula as \eqref{2AnU} except that $\t_j$ need
to be replaced by $\psi_j$ and the summation on $j$ starts 
from $j =0$. We refer to \cite{XTC} for the discussions of these 
two scanning geometries and their implementation in practical 
problems. 

Both types of these two OPED algorithms work well in our 
numerical testing (\cite{XT, XTC}).  It should be mentioned 
that the explicit formula of  $U_n(t)$ in \eqref{cheby} permits a 
fast implementation of the OPED algorithm, which uses fast 
Fourier sine transform and an interpolation step (\cite{XT}). 
\qed

\medskip\noindent
{\bf The case $d=3$.} In this case we can start from the formula
of $S_n f$ at \eqref{d3Sn}. We apply the Gaussian quadrature 
formula
$$
 \frac{3}{4} \int_{-1}^1f(t) (1-t^2) dt =
    \sum_{j=0}^n w_j f(t_j), 
$$
where $t_j$, $0 \le j \le n$, are zeors of $C_n^{3/2}(t)$. We can 
also apply the Gaussian quadrature formula for the Lebesgue
measure. This leads to a three dimensional OPED algorithm, 
\begin{equation} \label{A_3d}
 \CA_n f(x) =  \frac{1}{n+1} \sum_{k=0}^n \lambda_k
     \sum_{\nu=0}^{n} \sum_{j=0}^n w_j \CR f (\xi_{k,\nu}, t_j)
     \Phi_n(t_j, \langle  \xi_{k,\nu},x\rangle).
\end{equation}
The Radon data used in \eqref{A_3d} are integrals over 
planes $\langle x, \xi_{k,\nu}\rangle = t_j$. Such data can be
approximated by integrals over lines.  \qed

\medskip

For $d=3$, one can uses multiple 2D slices to reconstruct image
on a cylindrical domain, as proposed in \cite{X05}. An interesting
question is to see which of these two algorithms are more suitable
for the 3D reconstruction. 

Next we consider the convergence of $\CA_n f$ in the uniform
norm on $B^d$. 

\begin{thm}
The uniform norm of the operator $\CA_n$ is given by
\begin{equation}\label{A_norm}
    \|\CA\|_\infty = \sup_{x \in B^d} \Lambda_n(x), \quad 
     \Lambda_n(x) = 
   \sum_{\nu =1}^N \l_\nu \sum_{j=0}^n w_j (1-t_j^2)^{\frac{d-1}{2}} 
      \left |  \Phi_n(t_j, \langle x,\xi_\nu\rangle \right) |.  
\end{equation}
Furthermore, there is a constant $c$ independent of $n$, such that 
\begin{equation} \label{CAnNorm}
   \|\CA \|_\infty \le c\, n^{2d}. 
\end{equation}
In particular, if $f$ is smooth enough then $\CA_{2n}f$ converges to 
$f$ uniformly on $B^d$.
\end{thm}

\begin{proof}
To estimate the norm of $\CA_n$, we first observe
that 
$$
   \left |  (1-t^2)^{- \frac{d-1}{2}} \CR f(\xi_\nu, t)\right |
        \le b_{d-1} \|f\|_\infty
$$
from which it follows that 
$$
   \| \CA_n f \|_\infty \le  \|f\|_\infty
   \sum_{\nu =1}^N \l_\nu \sum_{j=0}^n 
            w_j (1-t_j^2)^{\frac{d-1}{2}} 
   \left | \Phi_n(t_j, \langle x,\xi_\nu\rangle) \right |, 
$$
since $b_{d-1} b_d^{-1} = c_{d/2}$. Taking the maximum over
$B^d$ shows that $\|\CA\|_\infty$ is bounded by the right hand
side of \eqref{A_norm}. To prove the equal sign, we construct
a function $f_\varepsilon$ for each $\varepsilon > 0$ such 
that $\|f_\varepsilon\|_\infty =1$ and $\|\CA f_\varepsilon\|_\infty
 \ge \max_{x\in B^d} \Lambda_n(x)- c \varepsilon$. A moment of 
reflection shows that the construction can be carried out easily; 
see \cite{X05} for one special case of $d =2$. 
 
To prove \eqref{CAnNorm} we use \eqref{eq:Phi} and the fact 
that $|C_n^\lambda(t)|
\le C_n^{\lambda}(1) = \binom{n+2\l -1}{n} = \CO(n^{2\l-1})$,
which implies that 
$$
 |\Phi_n(\xi,t)| \le  \sum_{k=0}^n \frac{k+d/2}{d/2} [C_n^{d/2}(1)]^2
    \le c \sum_{k=0}^n \frac{k+d/2}{d/2} k^{2d -2}
    \le c \, n^{2d}. 
$$
Since $\l_\mu$ and $w_j$ are all positive and, as the cubature 
and the quadrature are exactly for constant function,
$\sum_{\nu =1}^N \l_\nu =1$ and $ \sum_{j=0}^n w_j =1$, we
conclude that $\|\CA_n\| \le c\, n^{2d}$.  If $f \in C^{2d}$, then 
the fact that $\CA_n p = p$ for $p \in \Pi_n^d$ and the triangle 
inequality shows that 
$$
   \|\CA_n f- f\|_\infty \le (1 + \|\CA\|_\infty) E_n (f)_\infty
          \le c \,n^{2d} E_n (f)_\infty.
$$
It is shown in \cite{X05} that $E_n(f) \le c n^{-2r} \|\CD^r f\|$,
where $\CD$ is a second order differential operator, so that 
the convergence of $\CA_n f$ for functions smooth 
enough follows.
\end{proof}

We should point out that the estimate \eqref{CAnNorm} is 
a rough upper bound, the actual norm should be smaller. In 
fact, in the case of $d =2$, the norm of $\CA_{2m}$ at 
\eqref{2AnU} was estimated in \cite{X05} to be 
$$
   \|\CA_{2m}\|_\infty \sim m \log (m+1),
$$
which is sharp and is just slightly worse than the estimate 
\eqref{SnNorm} of the norm of the partial sum operator $S_n$
from which $\CA_{2m}$ is obtained. The proof of such a 
sharp estimate is rather involved and requires detail knowledge
of the zeros and weights of the quadrature and cubature
formulas. On the other hand, a result in \cite{Su} shows that 
the norm of any projection operator from $C(B^d)$ to $\Pi_n^d$ 
is at least $\CO(n^{\frac{d-1}{2}})$ for $d \ge 2$. As $\CA_n$ in 
\eqref{CAn} is in fact a projection operator, its norm cannot 
be bounded. We expect that the norm is in the order of 
$\CO(n^{d/2})$ multiplied by a log factor. 

It should be mentioned that other polynomial based algorithms
may have better approximation property (\cite{Ma, MN}. However, 
the polynomial preserving property seems to be an important
characteristic of OPED and using the partial sum allows also fast
implementation of the algorithm. The numerical tests show that
OPED works very well even for step functions such as 
Logan-Sheff head phantom \cite{XT,XTC}.


\section{Singular value decomposition of the Radon transform}
\setcounter{equation}{0}

Let $A: H \mapsto K$ be a linear continuous operator, where 
$H$ and $K$ are Hilbert spaces. Let $\{f_k\}_{k \ge 0}$ and 
$\{g_k\}_{k \ge 0}$ be orthonormal systems with respect to the 
inner product $\langle \cdot,\cdot\rangle_H$ in  $H$ and 
$\langle \cdot,\cdot \rangle_K$ in $K$, respectively. The singular 
value decomposition of $A$ is a representation
\begin{equation}\label{Af}
    A f = \sum_{k=1}^\infty \gamma_k \langle f,f_k\rangle_H g_k,
\end{equation}
where $\g_k$ are the singular values of $A$. Let $A^*$ be
the adjoint of $A$. Then
\begin{equation}\label{A*f}
   A^* g = \sum_{k=1}^\infty \g_k \langle g,g_k \rangle_K f_k. 
\end{equation}
Evidently $A f_k = \g_k g_k$ and $A^*g_k = \g_k f_k$.
Furthermore,  the generalized inverse of $A$ is given by
\begin{equation}\label{A+f}
  A^+ g = \sum_{k=0}^\infty  \g_k^{-1} \langle f,f_k \rangle_H  g_k.
\end{equation}

The singular value decomposition of the Radon transform 
was developed in \cite{Da, Lo} (see also \cite{N}). Let 
$Z = S^{d-1} \times [-1,1]$ and $w(t) = \sqrt{1-t^2}$, and
denote by $L^2(Z, w^{1-d})$ the space of Lebesgue integrable 
functions 
$$
 L^2(Z, w^{1-d}): = \{ g:  g(-\xi,-t) = g(\xi,t), \quad 
        \|g\|_{L^2(Z)} < \infty\}, 
$$   
where $\|g\|_{L^2(Z)}^2 =  \langle g, g \rangle_{L^2(Z)}$ and 
the inner product is defined by
$$
    \langle f, g \rangle_{L^2(Z)} : = c_{d/2} \int_{-1}^1 
           \sigma_d^{-1} \int_{S^{d-1}} f(\xi,t)g(\xi,t) d\omega(\xi) 
             (1- t^2)^{\frac{1-d}{2}} dt,
$$
in which $c_{d/2}$ is defined as in \eqref{Gegen}.  Then it is 
known (see, for example, \cite{N}) that 
$$
    \CR: L^2(B^d) \mapsto L^2 (Z, w^{1-d}) 
$$
is continuous. An orthonormal basis of $L^2(Z, w^{1-d})$ is 
readily available.

\begin{prop}
Let $\{Y_{j,m}: 1 \le j \le \dim \CH_m^d\}$ denote an 
orthogomal basis of $\CH_m^d$ and define 
\begin{equation} \label{gkj}
   g_{k,j}^n (\xi,t) =  [h_n^{(d/2)}]^{-1/2}
            (1-t^2)^{\frac{d-1}{2}} C_n^{d/2}(t) Y_{j,n-2k}(\xi),
\end{equation}
where $h_n^{(d/2)}$ is defined in \eqref{Gegen}. Then the functions 
$\{g_{k,j}^n: 0 \le 2k \le n, 1 \le j \le \dim\CH_{n-2k}^d\}$ forms an
orthogonomral basis for $L^2(Z,w^{1-d})$. 
\end{prop}

\begin{proof}
It is straightforward to verify that $\{g_{k,j}^n\}$ form an orthonormal 
system of  $L^2(Z,w^{1-d})$.  Let $g \in L^2(Z,w^{1-d})$. Then 
$w^{2d-2} g$ can be expanded in terms of the product orthonomal 
basis  $\{[h_n^{(d/2)}]^{-1/2} C_n^{d/2}(t) Y_{j, n-m}(\xi): 0 \le m \le n,
  0 \le j \le \dim\CH_{n-m}^d\}$ of $L^2(Z, w^{d-1})$. The condition 
 $g(-\xi, -t) = g(\xi,t)$ shows that the coefficients of the expansion 
are zero whenever $m$ is odd, so that we can assume 
$m = 2k$ and the expansion is uniquely determined.  
\end{proof}

Using $f_{k,j}^n$ in \eqref{OPbasis} and $g_{k,j}^n$ \eqref{gkj}, the 
singular value decomposition of the Radon transform at 
\eqref{Af}, \eqref{A*f} and \eqref{A+f} become the following: 

\begin{thm} 
Assume $f$ is in the Schwartz space. The singular decomposition of 
$\CR f$ is 
\begin{align} \label{Rf}
  \CR f  = \sum_{n=0}^\infty \g_n \sum_{0\le 2k \le n} 
        \sum_{j=0}^{M_{n-2k}}
      \langle f, f_{k,j}^n \rangle_{L^2(B^d)} g_{k,j}^n 
\end{align} 
where $M_m= \dim \CH_{m}^d$, $c_{d/2}$ is defined at  
\eqref{Gegen}; and 
\begin{align}\label{R*f}
  \CR^* g  = \sum_{n=0}^\infty \g_n \sum_{0\le 2k \le n} 
        \sum_{j=0}^{M_{n-2k}}
      \langle g, g_{k,j}^n \rangle_{L^2(Z)} f_{k,j}^n.  
\end{align} 
Furthermore, 
\begin{equation}\label{R+f}
   f(x) =  \sum_{n=0}^\infty \g_n^{-1} \sum_{0\le 2k \le n} 
        \sum_{j=0}^{M_{n-2k}} 
      \langle g, g_{k,j}^n \rangle_{L^2(Z)} f_{k,j}^n.
\end{equation} 
\end{thm}

These equations are the realization of \eqref{Af}, \eqref{A*f} and
\eqref{A+f} for the Radon transform. They are exactly the SVD 
derived in \cite{Da, Lo}, once the difference in notations is accounted 
for. 

Below we derive the singular value decomposition using our notation 
here. We need a proposition that goes back to \cite{Marr} when $d=2$. 

\begin{prop}
Let $P \in \CV_n^d$. Then for each $t \in [-1,1]$ and $\xi \in 
S^{d-1}$, 
\begin{equation} \label{R-OP}
   \CR P(\xi, t) = b_{d-1} (1-t^2)^{\frac{d-1}{2}} 
       \frac{C_n^{d/2}(t)}{C_n^{d/2}(1)} P(\xi). 
\end{equation}
In particular, the above formula applies to harmonic polynomials of 
degree $n$. 
\end{prop}

\begin{proof}
Let $Q_\xi$ be an orthogonal matrix whose first row is $\xi$. 
Then \eqref{Radon} shows that
$$
   \CR P(\xi,t) = (1-t^2)^{\frac{d-1}{2}} \int_{B^{d-1}} 
                            P( (t, \sqrt{1-t^2}\, y)Q_\xi)dy.  
$$      
The integral is a polynomial of $t$ since an odd power of 
$\sqrt{1-t^2}$ is always companioned by $y^\alpha$ with
$|\alpha|$ being odd, which has integral zero. Therefore,
$g(t) = (1-t^2)^{-\frac{d-1}{2}} \CR f(\xi,t)$ is of degree $k$
in $t$. Furthermore, the integral shows that 
$$
g(1) = \vol (B^{d-1}) P(\xi) =  b_{d-1} P(\xi).   
$$
If $g_j \in \Pi_j^d$ for $0 \le j \le n-1$, then the equation 
\eqref{IntRf} and the fact that $P \in \CV_n^d$ lead to 
$$
  \int_{-1}^1g(t)  g_j(t) (1-t^2)^{\frac{d-1}{2}} dt 
    = \int_{B^d} P(x) g_j(\langle x,\xi \rangle)dx =0,
$$
which shows immediately that the polynomial $g(t)$ is an
orthogonal polynomial with respect to $(1-t^2)^{\frac{d-1}{2}}$
on $[-1,1]$, that is, 
$$
    g(t) = (1-t^2)^{-\frac{d-1}{2}} \CR f(\xi,t) = a\, C_n^{d/2}(t).
$$
Setting $t =1$ determines the constant $a$ and completes the
proof. Finally, \eqref{OPbasis} with $k=0$ show that harmonic
polynomials of degree $n$ are in $\CV_n^d$.
\end{proof}

\begin{cor}
Let $f_{k,j}^n$ be the orthonormal basis of $\CV_n^d$ given
in \eqref{OPbasis}. Then
$$
  \CR f_{k,j}^n(\xi,t) = \gamma_n g_{k,j}^n(\xi,t),
$$
where the singular values $\gamma_n$ of $\CR f$ are 
given by
\begin{equation} \label{svd}
       \g_n =  b_{d-1} \sqrt{n!/(d)_n}.     
\end{equation}
\end{cor} 

\begin{proof}
Using \eqref{f_xi} and \eqref{gkj}, the equation \eqref{R-OP}
shows 
$$
 \CR f_{k,j}^n(\xi, t) = b_{d-1} [h_n^{(d/2)}]^{1/2}
    (1-t^2)^{\frac{d-1}{2}}
      \frac{C_n^{d/2}(t)}{C_n^{d/2}(1)} = \g_n  g_{k,j}^n(\xi,t), 
$$
where $\g_n = b_{d-1} [h_n^{(d/2)}]^{-1/2} H_n /C_n^{(d/2)}(1)$,
which can be simplified by using \eqref{Gegen}, \eqref{Hk} and 
the fact that $C_n^{(d/2)}(1) = (d)_n/n!$.
\end{proof}

\begin{thm} 
The singular decomposition of $\CR f$ satisfies
\begin{align} \label{Rf2}
  \CR f=c_{d/2} (1-s^2)^{\frac{d-1}{2}} 
             \sum_{n=0}^\infty  \left[h_n^{(d/2)}\right]^{-1} 
                     \int_{B^d} f(x) C_n^{d/2}(\langle x,\xi \rangle) dx \,
                  C_n^{d/2}(t),     
\end{align} 
where $M_m= \dim \CH_{m}^d$, $c_{d/2}$ is defined at  
\eqref{Gegen}; and 
\begin{align}\label{R*f2}
  \CR^* g = c\,
            \sum_{n=0}^\infty \g_n  \left[h_n^{(d/2)}\right]^{-1} 
                         \int_{-1}^1 \int_{S^{d-1}} \CR f(\xi,t) 
                     C_n^{d/2}(\langle x,\xi \rangle) C_n^{d/2}(t)
                       d\omega(\xi) dt.      
\end{align} 
where $c =  c_{d/2} b_{d-1} \sigma_d^{-1}$. 
\end{thm}

\begin{proof}
To prove \eqref{Rf2}, we note that by \eqref{f_xi},  
\begin{equation}\label{fg=ff}
 f_{k,j}^n(x) g_{k,j}^n(\xi, t) = [h_n^{(d/2)}]^{-1/2} H_n^{-1}
    (1-t^2)^{\frac{d-1}{2}} C_n^{d/2}(t) f_{k,j}^n(x) 
          f_{k,j}^n(\xi). 
\end{equation}
Since the constants are independent of $k$ and $j$, we can use
\eqref{reprod2} to write the summations in $k$ and $j$ of 
\eqref{Rf2} in a compact form. Collecting constants and using 
\eqref{Hk}, \eqref{svd} and \eqref{Gegen}, we easily verify that 
$$
   \g_n [h_n^{(d/2)}]^{-1/2}  H_n^{-1} f_{k,j}^n(x) \frac{n+d/2}{d/2}
     = b_{d-1}  [h_n^{(d/2)}]^{-1}. 
$$
Finally we note that $b_{d-1} b_d^{-1} = c_{d/2}$.  The proof of
\eqref{R*f2} is similar. 
\end{proof}

It is worth to comment that the two expressions \eqref{Rf2} and
\eqref{R*f2}
are independent of the choice of
orthonormal bases, and the equation \eqref{reprod2} implies 
that we can deduce the SVD from them using any 
orthonormal basis. In \cite{Da, Lo}, the SVD in terms of 
orthogonal basis with respect to $W_\mu$ is derived. In these 
more general cases, however, the simple analogue of the 
second equations of \eqref{Rf} and \eqref{R*f} do not hold. The 
reason again lies in the fact that the constant in \eqref{Hnk}
depends on $k$. 

Finally, by \eqref{A+f}, the truncation of the expansion of $f$ 
becomes 
\begin{equation*} 
   S_n^*f (x) =  \sum_{m=0}^n \g_m^{-1} \sum_{0\le 2k \le m} 
        \sum_{j=0}^{M_{m-2k}} 
      \langle g, g_{k,j}^m \rangle_{L^2(Z)} f_{k,j}^m.
\end{equation*}
Just as in the equations \eqref{Rf} and \eqref{R*f}, we can 
use \eqref{fg=ff} and \eqref{reprod2} to derive a compact
formula. The formula, however, is exactly $S_nf$. 
As a consequence, we see the truncated SVD algorithm
agrees with that formula \eqref{eq:Sn}. Hence, truncated SVD 
can be effectively implemented by using the OPED algorithm.

\end{document}